\documentclass[12pt, 
]{article}
\usepackage{booktabs}

\usepackage[utf8]{inputenc}
\usepackage{amssymb}
\usepackage{amsmath}

\usepackage{geometry}
\newtheorem{theorem}{Theorem}

\newtheorem{corollary}{Corollary}
\newtheorem{definition}{Definition}
\newtheorem{lemma}{Lemma}
\newtheorem{remark}{Remark}

\newenvironment{proof}[1][Proof]{\emph{#1.} }{\  \hfill $\square $ \vspace{5 pt}}

\usepackage{natbib}
\usepackage{amssymb}
\usepackage[dvips]{graphics}
\usepackage{amsmath}

\usepackage{mathpazo}
\usepackage{enumerate}

\usepackage{amsfonts}

\usepackage{amssymb}

\usepackage{enumerate}

\usepackage{mathrsfs}

\usepackage[usenames]{color}
\usepackage{cancel}

\usepackage{pgf,tikz}

\usetikzlibrary{decorations.pathreplacing,angles,quotes}
\usetikzlibrary{arrows.meta}
\usetikzlibrary{arrows, decorations.markings}
\tikzset{myptr/.style={decoration={markings,mark=at position 1 with %
       {\arrow[scale=2,>=stealth]{>}}},postaction={decorate}}}

\usepackage{hyperref}
\hypersetup{
    colorlinks,
    citecolor=blue,
    filecolor=black,
    linkcolor=blue,
    urlcolor=blue
}

\usepackage{pgf,tikz}
\usetikzlibrary{arrows}

\usepackage{fancyhdr}

\usepackage{soul}

\usepackage{emptypage}

\newcommand*\samethanks[1][\value{footnote}]{\footnotemark[#1]}

\usepackage[T1]{fontenc}
\DeclareFontFamily{T1}{calligra}{}
\DeclareFontShape{T1}{calligra}{m}{n}{<->s*[1.44]callig15}{}
\DeclareMathAlphabet\mathcalligra   {T1}{calligra} {m} {n}

\usepackage{multirow}
\usepackage{array}
\usepackage{mathtools}
\usepackage{arydshln}
\usepackage{threeparttable}
\usepackage{booktabs,caption}

\usepackage{ifthen}
\newboolean{showcomments}
\setboolean{showcomments}{true}

\newcommand{\pablo}[1]{  \ifthenelse{\boolean{showcomments}}
{\textcolor{green!50!black}{(T: #1)}}{}}
\newcommand{\marcelo}[1]{\ifthenelse{\boolean{showcomments}}
{\textcolor{red}{(M: #1)}}{}}
\newcommand{\agustin}[1]{  \ifthenelse{\boolean{showcomments}}
{\textcolor{blue!50!black}{(T: #1)}}{}}

\begin{document}

\title{Obvious Manipulations in Matching without and with
Contracts%
\thanks{%
We thank the participants of the 2024 Society for the Advancement of Economic Theory Conference for their valuable suggestions and comments. 
We acknowledge the financial support
from UNSL through grants 032016, 030120, from CONICET through grant
PIP 112-200801-00655, and from Agencia Nacional de Promoción Cient\'ifica y Tecnológica through grant PICT 2017-2355.}}


\author{R. Pablo Arribillaga\thanks{
Instituto de Matem\'atica Aplicada San Luis (CONICET-UNSL) and Departamento de Matemática, Universidad Nacional de San Luis, San Luis, Argentina, and RedNIE. Emails: \href{mailto:rarribi@unsl.edu.ar}{rarribi@unsl.edu.ar} 
and \href{mailto:ebpepa@unsl.edu.ar@unsl.edu.ar}{ebpepa@unsl.edu.ar}
} \and Eliana Pepa Risma\samethanks[2] 
}
\date{\today}

\maketitle
\begin{abstract}
This paper explores many-to-one matching models, both with and without contracts, where doctors' preferences are private and hospitals' preferences are public and substitutable. It is known that any stable-dominating mechanism --which is either stable or individually rational and Pareto-dominates (from the doctors' perspective) a stable mechanism--, is susceptible to manipulation by doctors. Our study focuses on \textit{obvious manipulations} and identifies stable-dominating mechanisms that prevent them. Without contracts, we show that more efficient mechanisms are less likely to be obviously manipulable and that any stable-dominating mechanism is not obviously manipulable. However, with contracts, none of these results hold. While we demonstrate that the Doctor-Proposing Deferred Acceptance (DA) Mechanism remains not obviously manipulable, we show that the Hospital-Proposing DA Mechanism and any efficient mechanism that Pareto-dominates the Doctor-Proposing DA Mechanism become (very) obviously manipulable, in the model with contracts.

\bigskip

 \bigskip


\end{abstract}

\section{Introduction}

A two-sided matching model with contracts involves a market composed of two distinct groups, typically referred to as doctors and hospitals. Each contract represents an agreement between a specific doctor and a specific hospital, and multiple contracts may exist between the same agents under different terms. The problem consists of assigning doctors to hospitals through contracts. Classical matching models (without contracts) can be seen as simplified cases where only one contract per doctor-hospital pair is possible. In many-to-one models, doctors can sign only one contract, while hospitals can sign multiple contracts. A set of contracts containing at most one contract per doctor is called an allocation and represents a potential outcome of the matching problem. All agents have preferences defined over allocations consisting of contracts that involve them. Two agents wishing to enter into an existing contract can do so freely, and they may also unilaterally terminate previous contracts if advantageous. An allocation is stable if no doctor-hospital pair desires to sign a contract outside the allocation, and no agent wishes to discard contracts already in it. \cite{hatfield2005matching} proved that if all hospitals have substitutable preferences,\footnote{Substitutability is a widely used condition in matching literature, meaning that hospitals do not consider contracts as complementary among themselves.} the set of stable allocations is nonempty. Furthermore, the most preferred stable allocations for each side of the market can be determined using generalizations of the Deferred Acceptance (DA) algorithms introduced by \cite{gale1962college}.

In our analysis, as is common in the literature, we assume that only one side of the market—the doctors—is strategic and possesses private information, while the preferences of hospitals are fixed and commonly known \citep[see, for example,][]{sakai2011note, hirata2017stable, iwase2022equivalence}. A (matching) mechanism is thus a function that, for each preference profile declared by the doctors, selects an allocation. A mechanism is considered stable if it always selects a stable allocation. Our focus will be on stable-dominating mechanisms, which are either stable mechanisms or individually rational mechanisms that Pareto-dominate (from the doctors' perspective) a stable mechanism. 

\cite{troyan2020obvious} pose the question: Why focus on stable-dominating mechanisms rather than exclusively on stable ones? The core issue is that stability is incompatible with Pareto efficiency. For example, the Doctor-proposing Deferred Acceptance Mechanism ($DDAM$), which Pareto dominates all other stable mechanisms, may still be Pareto inefficient. In the search for a normatively justified weakening of stability that remains compatible with efficiency, stable-dominating mechanisms have been examined in various studies \citep[see, for example,][]{kesten2010school, troyan2020essentially, hirata2017stable}.

In addition to $DDAM$, two other prominent stable-dominating mechanisms are the Hospital-proposing Deferred Acceptance Mechanism ($HDAM$) and the Efficiency-adjusted (Doctor-proposing) Deferred Acceptance Mechanism ($EADAM$) \citep{kesten2010school}, the latter of which has recently gained influence in market design by addressing the Pareto inefficiency of $DDAM$ \citep[see, for example,][]{cerrone2022school, chen2023regret, reny2022efficient, ortega2023cost}.

Besides stability and efficiency, the non-manipulability of matching mechanisms is crucial in the two-sided matching literature. Given that hospitals' preferences are public, we focus exclusively on manipulations that can be executed by doctors. A doctor is said to manipulate a mechanism if it can achieve a better outcome by misrepresenting its true preferences. It is well-established that in many-to-one matching models (with and without contracts) involving substitutable preferences, any stable-dominating mechanism may be susceptible to manipulation \citep[see][]{martinez2004group, hatfield2005matching, abdulkadirouglu2009strategy, kesten2010school, alva2019strategy}.

Given that manipulations cannot be completely avoided in this context, we seek stable-dominating mechanisms that at least prevent \emph{obvious manipulations}, as defined by \cite{troyan2020obvious}. A manipulation is considered 'obvious' if, in a specific and formal sense, it is significantly easier for agents to recognize and successfully execute compared to other types of manipulations. To define 'obvious,' we must specify how much information each agent has about the preferences of others. \cite{troyan2020obvious} assume that each agent operates under complete ignorance in this regard, focusing solely on the set of all possible outcomes that the mechanism could select based on its own report. A manipulation is deemed obvious if the best possible outcome under the manipulation is strictly better than the best possible outcome under truth-telling, or if the worst possible outcome under the manipulation is strictly better than the worst possible outcome under truth-telling. Furthermore, the term 'obvious' implies that an agent could infer that a mechanism is manipulable even without fully understanding its definition \citep[see Theorem 1 in][for this interpretation]{troyan2020obvious}. 

In the context of college admission, \cite{troyan2020obvious} prove that any stable-dominating mechanism is not obviously manipulable. In a context without contracts, we extend this result showing that it also holds for substitutable preferences. Our proof uses an original argument that follows the sequence a clear sequence.  First, we demonstrate that a mechanism which Pareto-dominates a stable and not obviously manipulable mechanism is itself not obviously manipulable. Second, we prove that $HDAM$ is not obviously manipulable. Finally, as a consequence of these two findings, we conclude that any stable-dominating mechanism is also not obviously manipulable. 

For the model with contracts, we prove that $DDAM$ remains not obviously manipulable. This result extends the findings of \cite{troyan2020obvious} to the model with contracts, although its proof requires entirely different arguments and techniques. Surprisingly, we also demonstrate that none of the three findings mentioned previously for the setting without contracts hold in the context with contracts, even in the one-to-one model.  
In fact, both $HDAM$ and any efficient mechanism that Pareto-dominates $DDAM$ (including $EADAM$) become \textit{(very) obviously manipulable} when contracts are introduced, even in the one-to-one case. We use the term \textit{``very''} to indicate that these manipulations are profitable under \textit{any} preference profile of the other doctors. Such manipulations appear perfectly natural and are likely to be recognized by agents. Therefore, we will refer to them as \textit{very obvious manipulations}, in line with the taxonomy of \cite{troyan2020obvious}. These manipulations not only make intuitive sense but also yield clear predictions that could be tested in laboratory experiments. This contrast highlights a significant difference between models with and without contracts regarding the strategic behavior of agents.

\cite{troyan2020essentially} apply the notion of obvious manipulation to one-sided matching markets, introducing the concept of essential stable matching and demonstrating that no essentially stable mechanism is obviously manipulable. Other recent papers exploring the concept of obvious manipulation in contexts beyond two-sided matching markets include \cite{aziz2021obvious} and \cite{arribillaga2024obvious} in the context of voting, \cite{ortega2022obvious} in cake-cutting, and \cite{psomas2022fair} in allocation problems.

The rest of the paper is organized as follows. Section \ref{section prelim} introduces the model and the concepts of stable-dominating mechanisms and obvious manipulations. Section \ref{section main} presents the main results of our study in two subsections: Subsection \ref{with} covers the results obtained in matching without contracts, while Subsection \ref{without} discusses the results in matching with contracts. Finally, Section \ref{section final} offers concluding remarks. 

\section{Preliminaries}\label{section prelim}
\subsection{Matching model with contracts and strategic doctors}\label{model}

We consider a many-to-one matching model with contracts, where the market consists of two disjoint sides: a finite set of doctors $D$ and a finite set of hospitals $H$. The problem involves assigning agents from one side of the market to agents on the opposite side. However, unlike classical matching models without contracts, the contractual conditions (such as salary, schedules, work tasks, etc.) that characterize the relationship between two agents are not predetermined. In a particular market, there is a finite universal set of contracts $\mathbf{X}$. Each contract $x \in \mathbf{X}$ is bilateral, involving exactly one doctor $x_{D} \in D$ and one hospital $x_{H} \in H$. The set $\mathbf{X}$ may include two or more contracts involving the same pair of agents $(d,h) \in \mathbf{D \times H}$, but under different conditions. The classical matching model without contracts can be seen as a special case of this setting, where $\mathbf{X}$ contains exactly one contract involving each pair of agents $(d,h) \in \mathbf{D \times H}$.

In the many-to-one matching model that we study, each hospital can sign multiple contracts, while each doctor can sign at most one contract. Given this requirement, the possible solutions to the assignment problem, referred to as allocations, can be characterized as follows: a set of contracts $Z\subseteq \mathbf{X}$ is an \textbf{%
allocation}\ if $x\neq y$ implies $x_{D}\neq y_{D}$ for all $x,y\in Z$. 
For a given set of contracts $Y \subseteq \mathbf{X}$ and an agent $i \in D \cup H$, we denote by $A(Y)$ the set of all allocations that are subsets of $Y$, and by $Y_{i}$ the set of all contracts in $Y$ involving $i$. Note that the empty set (representing a situation where no contracts are signed) is also an allocation, and $\varnothing \in A(Y)$ for all $Y \subseteq \mathbf{X}$.

Given a set of contracts $\mathbf{X}$, a particular market is defined by a preference relation for each agent $i \in D \cup H$ over the set of allocations $A(\mathbf{X}{i})$.\footnote{Preferences are antisymmetric, transitive, and complete binary relations on $A(\mathbf{X}{i})$. Note that every allocation in $A(\mathbf{X}_{i})$ contains only contracts involving $i$.} In our analysis, as is common in the literature, we assume that only one side of the market is strategic—the doctors—while hospitals' preferences are fixed and commonly known \citep[see, for example][]{sakai2011note, hirata2017stable, iwase2022equivalence}.

Observe that $\left\vert Z\right\vert \leq 1$ for all $Z\in A(\mathbf{X}{d})$ and $d\in D$. Therefore, we can visualize a doctor's preference relation as an order over the contracts involving that doctor, along with the option of the empty set. An arbitrary preference for doctor $d$ will be denoted by $P{d}$, and the associated weak preference relation will be represented by $R_{d}$.\footnote{That is, for all $x, y \in A(\mathbf{X}{i})$, we have $xR{d}y$ if and only if either $x = y$ or $xP_{d}y$.} The set of all possible preference relations for a doctor $d$ in a given market will be denoted by $\mathcal{P}{d}$. A preference profile $P = (P{d}){d \in D}$ specifies a preference relation for each doctor, and the set of all possible preference profiles in the market is represented by $\mathcal{P} = \prod{d \in D} \mathcal{P}{d}$. Finally, for each profile $P$ and doctor $d \in D$, we will denote by $P{-d}$ the subprofile in $\mathcal{P}{-d} = \prod{i \in D \setminus {d}} \mathcal{P}{i}$ obtained by removing $P{d}$ from $P$.

\begin{definition}
A \textbf{(matching) mechanism} is a function $\phi :\mathcal{P}%
\mathbf{\rightarrow }A(\mathbf{X})$ that returns an allocation in $A(\mathbf{%
X})$ for each profile of preferences $P\in \mathcal{P}.$\medskip
\end{definition}

Given $d\in D,$ we denote by $\phi _{d}(P)$ the (unique) contract in $%
\phi (P)$ involving $d$, if such a contract exists. Otherwise,  $\phi _{d}(P)=\emptyset$.

\subsection{Stability and Efficiency}

An essential property in matching models is the stability of allocations. Unstable outcomes may not persist over time due to agents' freedom to cancel or sign contracts at any moment. Agents are expected to act in their best interest by unilaterally terminating some contracts or entering into new ones, provided the respective other parties are also willing to sign. 
To introduce the notion of stability, it is necessary to consider the preferences of hospitals over subsets of contracts. As previously mentioned, each hospital $h\in H$ has an
antisymmetric, transitive and complete preference relation over $%
A ( \mathbf{X}_{h}) $, denoted as $\succ _{h}.$ \footnote{A
preference profile for hospitals will be denoted by $\succ :=(\succ _{h})_{h\in
H} $.} The \textbf{choice set} of $h\in H$ given  $Y\subseteq \mathbf{X,}$ is the subset of $Y_{h}$ that $h$ prefers most
according to $\succ _{h}$. Formally, we define this choice set as $C_{h} ( \succ _{h},Y) =\max_{\succ _{h}}A(Y_{h})$. Similarly, the choice set for a doctor $d\in D$ given a preference $%
P_{d}\in \mathcal{P}_d$ and $Y\subseteq \mathbf{X,}$  is $C_{d} ( P_{d},Y) =\max_{P_{d}}A(Y_{d}).$\footnote{%
Note that for each $d\in D,$ the choice set $C_{d} ( P_{d},Y) $
includes only its best acceptable contract in $Y_{d}$ according to $P_{d}$, if any.} 
Note that a choice set could be the empty set. As is customary in the matching literature, we will assume that hospitals' preferences are substitutable. This means that hospitals do not view contracts as complementary to one another.  Formally, $\succ _{h}$ satisfies \textbf{%
substitutability} if $x\in C_{h} ( \succ _{h},W) $ implies $x\in
C_{h} ( \succ _{h},Y) $ whenever $x\in Y_{h}\subseteq 
W_{h}\subseteq \mathbf{X}.$ Throughout the paper, we will assume that $\succ_{h}$ is substitutable, fixed, and known to doctors, for each hospital $h$. To simplify notation, we will omit preferences in the choice set notation whenever they are clear from the context. Thus, we will write $C_{h} ( Y) $ and $%
C_{d} ( Y) $ instead of $C_{h} (\succ _{h},Y) $ and $%
C_{d} ( P_{d},Y)$, respectively.  Furthermore, for $Y\subseteq \mathbf{X}$ we will denote $C_{H} ( Y)
=\cup_{h\in H}C_{h} ( Y) $ and $C_{D} ( Y)
=\cup_{d\in D}C_{d} ( Y) .$ 

Certain well-known properties of choice sets of substitutable preferences almost trivially follow from their definition. We mention them here because they will be used to show our results.
For all $X,Y\subseteq \mathbf{X}$ and $i\in D\cup H$: 
\textit{(i)} $C_{i} ( Y) \subseteq Y$;
\textit{(ii)} $C_{i} ( Y) \subseteq X\subseteq Y$ implies $C_{i} (X) =C_{i} ( Y) $; 
\textit{(iii)} $C_{i} ( C_{i} ( Y) ) =C_{i} ( Y)$; 
\textit{(iv)} If $i$'s preferences satisfy substitutability, then $C_{i} ( X\cup
Y) =$ $C_{i} ( C_{i} ( X) \cup Y).$ 

Let $Y\in A(\mathbf{X})$ be an allocation and $P\in \mathcal{P}$.  $Y$ is \textbf{individually rational} if  $C_{D} ( Y) =C_{H} ( Y) =Y$, meaning it does not include any unwanted contracts. The allocation $Y$ is \textbf{blocked} by a contract $x\in \mathbf{X}\setminus Y$ if $x\in C_{x_{D}} ( Y\cup  \{ x\} ) \cap C_{x_{H}} (Y\cup  \{ x\} )$; that is, if both involved agents prefer to add the contract $x$ to $Y$ and may subsequently choose to cancel some of their original contracts in $Y$.  The allocation $Y$ is \textbf{stable} if it is individually rational and cannot be blocked by any contract.

The allocation $Y\in A(\mathbf{X})$ is said to \textbf{Pareto-dominates} another allocation $Y'\in A(\mathbf{X})$ if 
$Y_d R_d Y'_d$ for all $d\in D$ and $Y_d P_d Y'_d$ for some $d\in D$.  An allocation is said to be  \textbf{stable-dominating} if it is either stable or individually rational and Pareto-dominates some stable allocation. Additionally, an allocation is termed \textbf{efficient}  if there is not another allocation that Pareto-dominates it.

\begin{definition}
A mechanism $\phi :\mathcal{P}\mathbf{\rightarrow }A(\mathbf{X})$  is \textbf{stable-dominating} if for all $P\mathcal{\in P}$, the allocation $\phi
(P)$ is stable-dominating at $P$. A mechanism $\phi :\mathcal{P}\mathbf{\rightarrow }A(\mathbf{X})$ is \textbf{efficient} if for all $P\mathcal{\in P}$, the allocation $\phi
(P)$ is efficient at $P$.
\end{definition}

\cite{hatfield2005matching} show that if all hospitals have substitutable preferences, then there exists a unanimously most preferred stable allocation for each market side, which can be obtained using both the doctor-proposing and hospital-proposing Deferred Acceptance algorithms. Consequently, the set of stable allocations is nonempty.

Before concluding this subsection, we formally introduce three prominent stable-dominating mechanisms: the Doctor-proposing Deferred Acceptance Mechanism and the Hospital-proposing Deferred Acceptance Mechanism, both of which are classical and highly relevant in the literature; as well as the Efficiency-adjusted Doctor-proposing Deferred Acceptance Mechanism \citep{kesten2010school}, which has recently gained prominence in market design by correcting the Pareto inefficiencies inherent in the Doctor-proposing Deferred Acceptance Mechanism.
The Doctor-proposing Deferred Acceptance Mechanism can be computed using a deferred acceptance algorithm where doctors make offers: Once a profile $P\in \mathcal{P}$ is fixed, each doctor proposes its most preferred contract that has not been rejected in previous steps, while each hospital accepts contracts from the set of accumulated offers received. The algorithm terminates when all offers are accepted or no more offers remain. The output of the algorithm is the set of contracts (allocation) accepted by the hospitals in the final iteration. We now present the formal definition of the mentioned algorithm. \newline

\textbf{The Doctor-proposing Deferred Acceptance Algorithm (DDA algorithm)}

\noindent \textit{Input:} \newline
A market $(\mathbf{X},P).$\newline
\textit{Begin:}\newline
Set $X^{1}=\mathbf{X}$ and $t:=1.$\newline
\textit{Repeat}:

\textit{Step 1}: Determine the set of contracts that doctors offer in the
iteration $t,$ this is, \newline$O^{t}:=C_{D} ( X^{t}) .$

\textit{Step 2}: From the set of accumulated offers, $O_A^t:=\cup_{k=1}^{t}O^{k}$, determine $C_{H} ( O_A^t) .$ This is the set of contracts (provisionally) accepted by hospitals in the iteration $t$.  \newline
\ \ \ \ \ \ \ \ \ \ \ If $C_{H} ( O_A^{t}) =O^{t}$, the algorithm
stops with output $C_{H} ( O_A^{t}) .$\newline
\ \ \ \ \ \ \ \ \ \ \ If $C_{H} ( O_A^{t}) \neq O^{t}$,
define $X^{t+1}:=X^{t}\setminus(O^t\setminus C_H(O_A^t))$,
 this is, the set of contracts that have not been rejected yet; set $t:=t+1$; and repeat Steps 1 and 2.\newline
\textit{End}\medskip

Let $DDAM:\mathcal{P}\mathbf{\rightarrow }A(\mathbf{X})$ be
the \textbf{Doctor-proposing Deferred Acceptance Mechanism}, which returns the stable allocation obtained through the DDA algorithm for each preference profile $P\mathcal{\in P}$.

Symmetrically, we con define the \textbf{Hospital-proposing Deferred Acceptance Algorithm } (HDA algorithm) by interchanging the roles of $D$ and $H$.
Let $HDAM:\mathcal{P}\mathbf{\rightarrow }A(\mathbf{X})$ be
the \textbf{Hospital-proposing Deferred Acceptance Mechanism}, which returns the stable allocation obtained through the HDA algorithm for each preference profile $P\mathcal{\in P}$. 

To introduce the Efficiency-adjusted Doctor-proposing Mechanism, we follow the original framework outlined by \cite{kesten2010school} and adapt it to a context with contracts. This mechanism relies on the idea of identifying interrupting pairs (and/or contracts) in the  DDA algorithm and canceling the applications of the interrupting doctors to the corresponding critical hospitals. 
Consider a scenario where the DDA algorithm is applied. Let $d$ be a doctor who is tentatively placed at a hospital  $h$ with contracts $x$ at some Step $t$ and is subsequently rejected from it at a later Step $t'$. If there is at least one other doctor (with some contract) who is rejected by hospital $h$ after Step $t-1$ and before Step $t'$ -- that is, rejected at a Step $l\in\{t, t+1, . . . , t'-1\}$ --, then $x$ is termed an \textit{interrupting contract} of Step $t'$.\footnote{In the context of \cite{kesten2010school}, we refer to $d$ as an \textit{interrupter} for hospital $h$ and the pair $(d,h)$ if an \textit{interrupting pair}  of Step $t'$. However, in our context, the contract involved in the interruption must also be taken into account. As far as we know, this work is the first to present a version of the EADAM in a context involving contracts.} \newline

\textbf{The Efficiency-Adjusted (Doctor-proposing) Deferred
Acceptance Algorithm}
\noindent \newline  
\textit{Input:} \newline
A market $(\mathbf{X},P).$

\textit{Round $0$}:\newline Run the DDA algorithm.
 
\textit{Round $k$, $k \geq 1$}: \newline   Find the last step of the DDA algorithm executed in Round 
$(k-1)$ where an interrupting contract was rejected by the respective hospital. Identify all interrupting contracts at that step.

If no interrupting contracts are found, terminate the process and consider the allocation obtained in Round 
$(k-1)$ as the final output.

Otherwise, for each identified interrupting contract $x$, remove it from the preference list of the corresponding doctor 
$x_D$, while preserving the relative order of the remaining contracts. After updating the preference profile, rerun the DDA algorithm and proceed to a new round.\footnote{For simplicity, we assume that every doctor consents to modify their preferences.}

\textit{End}\medskip

Let $EADAM:\mathcal{P}\mathbf{\rightarrow }A(\mathbf{X})$ be
the \textbf{Efficiency-Adjusted (Doctor-proposing) Deferred
Acceptance Mechanism},  which returns the stable allocation obtained by the Efficiency-Adjusted (Doctor-proposing) Deferred Acceptance  Algorithm for each preference profile $P\mathcal{\in P}$.

Given two mechanisms $\phi:\mathcal{P}\mathbf{\rightarrow }A(\mathbf{X})$ and $\phi':\mathcal{P}\mathbf{\rightarrow }A(\mathbf{X})$, we say that $\phi'$ \textbf{Pareto-dominates} $\phi$ if $\phi'\neq\phi$ and $\phi'_d(P)R_d\phi_d(P)$, for each $d\in D$ and each $P\in \mathcal{P}$. It is known that any stable-dominating mechanism Pareto-dominates $HDAM$; we state this as a remark for future reference. 

\begin{remark}\label{duality}
Any stable-dominating  mechanism $\phi':\mathcal{P}\mathbf{\rightarrow }A(\mathbf{X})$ such that $\phi'\neq HDAM$  Pareto-dominates $HDAM$. 
\end{remark}

\subsection{Manipulations and Obvious Manipulations}

The concept of non-manipulability, also known as strategy-proofness, has been central to studying agents' strategic behavior. A doctor is said to manipulate a matching mechanism if a situation exists where it achieves a better outcome by reporting a preference different from its true one.

\begin{definition}
Given a mechanism $\phi :$ $\mathcal{P}\mathbf{\rightarrow }A(\mathbf{X})$ and $d\in D$ with true preference $P_{d}\in \mathcal{P}_{d}$, the preference $P_{d}^{\prime }\in \mathcal{P}_{d}$ is a \textbf{manipulation} of $\phi $ at $P_{d}$ if there is a (sub)profile $P_{-d}\in \mathcal{P}%
_{-d}$ such that 
\begin{equation}
\phi _{d}(P_{d}^{\prime },P_{-d})\text{ }P_{d}\text{ }\phi
_{d}(P_{d},P_{-d}).
\end{equation}%
A mechanism is \textbf{non-manipulable} if no manipulation is possible.
\end{definition}

In the context of many-to-one matching markets (both with and without contracts) where hospitals have substitutable preferences, any stable mechanism is susceptible to manipulation \citep[see][]{martinez2004group,hatfield2005matching}.\footnote{According to Hatfield and Milgrom (2005), when substitutability is combined with a property known as the Law of Aggregate Demand, it ensures strategy-proofness. However, if only substitutability is assumed, this result does not hold.} The same applies to stable-dominating mechanisms \citep[see][]{alva2019strategy}. Therefore, our goal is to identify stable-dominating mechanisms that, at a minimum, avoid obvious manipulations, as defined by \cite{troyan2020obvious}. Intuitively, a manipulation is considered ``obvious'' if it makes the agent strictly better off either in the worst-case or best-case scenario.

In many-to-one matching markets (with or without contracts) with substitutable preferences, any stable mechanism is vulnerable to manipulation \citep[see][]{martinez2004group,hatfield2005matching}.\footnote{According to \cite{hatfield2005matching}, when substitutability is combined with a property known as the Law of Aggregate Demand, it ensures strategy-proofness. However, if only substitutability is assumed, this result does not hold.} The same applies to stable-dominating mechanisms \citep[see][]{alva2019strategy}. Therefore, we seek stable-dominating mechanisms that, at a minimum, avoid obvious manipulations as defined by \cite{troyan2020obvious}. Intuitively, a manipulation is considered ``obvious'' if it makes the agent strictly better off either in the worst-case or best-case scenario.

Given a mechanism $\phi :\mathcal{P}\mathbf{\rightarrow }A(\mathbf{X}),$ a doctor $%
d\in D$ and a preference $P_{d}\in \mathcal{P}_{d}$, we define the \textbf{%
option set} left open by $P_{d}$ at $\phi $ as 
\begin{equation*}
O^{\phi }(P_{d})=\{\phi _{d}(P_{d},P_{-d}):P_{-d}\in \mathcal{P}_{-d}\}.
\end{equation*}

Given $Y \subseteq \mathbf{X}$, let $W_{d}(P_{d},Y)$ represent the worst allocation in $Y_d$ according to preference $P_{d}$. Note that the best allocation in $Y_d$ according to $P_{d}$ is the same as the choice set $C_{d}(P_{d},Y)$.

\begin{definition}\label{defnom}
Let $\phi :\mathcal{P}\mathbf{\rightarrow }A(\mathbf{X})$ be a mechanism,  $%
d\in D$ with true preference  $P_{d}\in \mathcal{P}_{d}$, and let $P_{d}^{\prime }\in \mathcal{%
P}_{d}$ be a manipulation of $\phi $ at $P_{d}$. Then, $P_{d}^{\prime
}$ is an \textbf{obvious manipulation} if 
\begin{equation} 
W_{d}(P_{d},O^{\phi }(P_{d}^{\prime }))\ P_{d}\ W_{d}(P_{d},O^{\phi
}(P_{d})).  \label{1}
\end{equation}%
or 
\begin{equation}
C_{d} ( P_{d},O^{\phi }(P_{d}^{\prime })) \ P_{d}\ C_{d} (
P_{d},O^{\phi }(P_{d})) .  \label{11}
\end{equation}%
Mechanism $\phi$ is \textbf{not obviously manipulable (NOM)} if it does not admit obvious manipulations. 
Otherwise, $\phi$ is \textbf{obviously manipulable}.
\end{definition}

We conclude this section with two remarks concerning Definition \ref{defnom}. The first remark highlights that stable-dominating mechanisms do not permit obvious manipulations in the sense described by condition (\ref{11}). The reasoning is straightforward and applies uniformly across all settings discussed in this paper: the best outcome a doctor $d\in D$ can obtain from any stable-dominating mechanism is its most preferred contract among those acceptable to the corresponding hospitals if such a contract exists. This outcome cannot be improved by any deviation of $d$.

In fact, given $d\in D$ and $P_{d}\in \mathcal{P}_{d}$, define  
$\overline{X}= C_d(P_d, \{x\in \mathbf{X}_{d}:\{x\}\succ _{x_{H}}\varnothing \}).$  
Let  $\overline{P}_{-d}\in \mathcal{P}_{-d}$
be such that $\overline{P}_{i}=\varnothing $ for all $i\in D\setminus \{
d\} .$ Then,  $ 
\overline{X}$ is the only stable-dominating allocation in $( P_{d},\overline{P}%
_{-d})$ and, consequently, ${\phi }_{d} ( P_{d},\overline{P}
_{-d})=\overline{X}$ . This implies $C_{d} ( P_{d},O^{\phi }
(P_{d})) $ $R_{d}$ $\overline{X} .$
Furthermore, for any $P_{d}^{\prime }\in \mathcal{P}_{d}\setminus \{P_{d}\} $,  we have $O^{\phi }(
P_{d}^{\prime })\subseteq\{x\in\mathbf{X}_{d}:\{x\}\succ_{x_{_{H}}}\varnothing \}$ because $\phi $ is individually rational.\footnote{Note that no contract $z\in \mathbf{X}_{d}$ such that 
$ \varnothing \succ_{z_{_{H}}}\{z\}$ can be included in any individually rational allocation. 
}  Therefore, $\overline{X} $ $R_{d}$ $C_{d}(
P_{d},O^{\phi }(P_{d}^{\prime })) .$ Thus, 
$C_{d}( P_{d},O^{\phi }(P_{d})) \text{ }R_{d}\text{ }%
\overline{X}\text{ }R_{d}\text{ }C_{d}( P_{d},O^{%
\phi }(P_{d}^{\prime })).$

\begin{remark} \label{bestcase}
Assume that $\phi :\mathcal{P}\mathbf{\rightarrow }A(\mathbf{X})$ is a stable-dominating mechanism. 
Then, $\phi $ does not admit obvious manipulations in the sense of (\ref{11}).
\end{remark}

The second remark addresses the consideration of preference $P_{d}^{\prime }$ as a manipulation of preference $P_{d}$ in Definition  \ref{defnom}. This condition is not strictly necessary because we could impose requirement (\ref{1}) on any arbitrary preference, and the resulting definition would remain unchanged. The main argument for this observation is that, given that the set of feasible contracts $\mathbf{X}$ is finite, a preference $P_{d}^{\prime }$ that violates (\ref{1}) must necessarily be a manipulation of $P_{d}$. The profile $P_{-d}$ where the manipulation occurs is such that $W_{d}(P_{d},O^{\phi}(P_{d}))=\phi_d(P_{d},P_{-d})$.

\begin{remark}\label{equi}
A mechanism $\phi :\mathcal{P}\mathbf{\rightarrow }A(\mathbf{X})$ is not obviously manipulable
(NOM) if and only if
$$W_{d}(P_{d},O^{\phi }(P_{d}))\ R_{d}\ W_{d}(P_{d},O^{\phi
}(P_{d}^{\prime })).$$
for all $d\in D$ and all $P_{d},P_{d}^{\prime }\in \mathcal{P}_{d}$.
\end{remark}

\section{Main Results}\label{section main}
In this section, we present the core findings of the paper, structured into two subsections based on whether the results pertain to models without contracts or with contracts. Throughout all statements, we assume that all hospitals have substitutable preferences.

\subsection{ Matching without contracts} \label{with}

As previously mentioned, classical matching models without contracts can be viewed as special cases of settings with contracts by assuming that $\mathbf{X}$ contains at most one contract for each pair  $(d,h)\in \mathbf{D\times H}$. We first focus on such a model and demonstrate that any stable-dominating mechanism is NOM. This result extends the findings of \cite{troyan2020obvious} from responsive to substitutable preferences. We present a straightforward proof based on different arguments and techniques from those used in that paper. \cite{troyan2020obvious} begin by noting that all stable-dominating mechanisms yield the same worst-case outcome for a truth-telling doctor. They then provide a precise characterization of the worst possible outcome under a stable-dominating mechanism, given a doctor’s true preferences. To achieve this, they use Hall's Theorem \citep{https://doi.org/10.1112/jlms/s1-10.37.26}, which provides a necessary and sufficient condition for finding a matching that covers a bipartite graph. Finally, they examine particular manipulative reports that have received significant attention in the literature and demonstrate that neither constitutes an obvious manipulation under a stable-dominating mechanism. The rest of the proof follows naturally from these observations.

Our result is based on two findings. First, we show that any individual rational mechanism that Pareto-dominates a stable and NOM mechanism is also NOM. In some sense, this finding suggests that a more efficient mechanism has a lower likelihood of being obviously manipulable. Second, we examine $HDAM$ and prove that it is NOM.  Finally, we conclude that all stable-dominating mechanisms are NOM, as they are individually rational and Pareto-dominate $HDAM$.

To prove that an arbitrary individual rational mechanism,  $\phi'$, which Pareto-dominates a stable and NOM mechanism, $\phi$, is also NOM, we use the following ideas. 
First, since $\phi'$ Pareto-dominates $\phi$, we observe that the worst outcome a doctor can receive under  $\phi'$ when truthfully reporting its preference must be at least as good as the worst outcome under $\phi$ for the same true preference.
Second, consider an arbitrary doctor $d$ a possible deviation $P'_d$ (a misreport of preference).  For this deviation, we identify a suitable sub-profile $\widetilde{P}_{-d}$ of the remaining doctors such that: (i) under mechanism $\phi$, the worst possible outcome at truth-telling is at least as good as the outcome obtained with $(P'_d,\widetilde{P}_{-d})$; and (ii) for this sub-profile and the deviation $P'_d$, there is only one stable-dominating allocation and $\phi'$ and  $\phi$ must coincide, returning the same allocation given the specified profile.

Therefore, based on (i) and (ii), the worst outcome for doctor $d$ under $\phi$ at truth-telling is at least as good as the outcome $d$ would receive under $\phi$ or $\phi'$ when deviating to $P'_d$ while the remaining doctors declare the previously mentioned sub-profile. This outcome is, in turn, clearly at least as good as the worst outcome under $\phi'$ at $P'_d.$ Then, based on the first observation, the worst outcome that $d$ can obtain by declaring its true preference under $\phi'$ is at least as good as the worst outcome that $d$ can obtain by declaring $P'_d$ under $\phi'$.  

\begin{theorem}
\label{dominate NOM} Assume that for each pair $(d,h)\in D\times H$ there
is at most one contract $x\in \mathbf{X}$ such that $x_{D}=d$ and $x_{H}=h$, and that all agents have substitutable preferences.  Let $\phi :\mathcal{P}\mathbf{\rightarrow }A(\mathbf{X})$ be a stable and NOM mechanism. 
If $\phi' :\mathcal{P}\mathbf{\rightarrow }A(\mathbf{X})$ is an individual rational mechanism that Pareto-dominates $\phi$, then $\phi'$ is NOM.
\end{theorem}

\begin{proof} 
Given $d\in D$ and $P_{d}\in \mathcal{P}_{d}$, since $\phi'$ Pareto-dominates $\phi$ we have
\begin{equation} \label{ww}
    W(P_{d},O^{\phi' }(P_{d}))R_{d}W(P_{d},O^{\phi }(P_{d}))
\end{equation}

Furthermore, because $\phi$ is NOM and Remark \ref{equi}, for any arbitrary   $P_{d}^{\prime }\in \mathcal{P}_{d}$, there exists a profile $P_{-d}\in \mathcal{P}_{-d}$ such that
\begin{equation} \label{ww1}
W(P_{d},O^{\phi }(P_{d}))R_{d}\phi_d(P_d',{P}_{-d})
\end{equation}

We denote $%
X:=\phi( P'_{d},P_{-d})$ and consider the profile   $\widetilde{P}_{-d}\in \mathcal{P}_{-d}$ defined as follows:

$\widetilde{P}_{i}:=X%
_{i},\varnothing .$   for each $i\in
D\backslash \{ d\}$
($\widetilde{P}_{i}=\emptyset $  in case $X _{i}=\varnothing $)

Next, we state a claim related to this profile, which will be useful for proving the theorem. 

\underline{Claim:} $ \phi(P_d',P_{-d})=\phi(P_d',\widetilde{P}_{-d})=\phi'(P_d',\widetilde{P}_{-d}).$
Let $\widetilde{P}=(P_d',\widetilde{P}_{-d})$. We will show that $X$ is the only stable allocation under $\widetilde{P}$, immediately implying the first equality.
From the definition of $\widetilde{P}_{i}$   for each $i\in D\setminus \{ d\}$ and the fact that $X$ is stable under $(P'_d,P_{-d})$, it follows that $X$ is also stable under $\widetilde{P}$. 
Now, assume that there is another allocation $Y\subset \mathbf{X}$ such that $Y\neq X$ and $Y$ is stable under $\widetilde{P}$. 
By the definition of $\widetilde{P}_{-d}$  and the individual rationality of $Y$, we know that:
\begin{equation} \label{aaa}
    Y_i\subseteq X_i 
 \text{ for all }  i\in D\setminus \{d\}
\end{equation}
Moreover, $Y_d\setminus X_d\neq\varnothing$, otherwise $Y\subsetneq X$, and considering any contract  $w\in X\setminus Y$, would lead to a blocking contract for $Y$. Indeed, $w\in C_{w_D}(X)\cap C_{w_H}(X)$ because $X$ is individually rational under $\widetilde{P}$. Consequently, due to substitutability, $w\in C_{w_D}(Y\cup\{w\})\cap C_{w_H}(Y\cup\{w\})$.
 
Suppose that $Y_d=\{y\}$ for some contract $y$. We will prove that \begin{equation} \label{cc}
    \{y\} 
    P'_d  X_d 
\end{equation}

In the case where  $X_d=\varnothing$,  statement \ref{cc} follows trivially because $Y$ is individually rational. 
Assume now that  $X_d=\{x\}$,  then $x\neq y$ and we will assume, for the sake of contradiction, that $\{x\} P'_d  \{y\}$. Since there are not two different contracts involving the same doctor-hospital pair, it must be  $y_H\neq x_H$. Therefore, all contracts in $Y_{x_H}$ involve doctors in $D\setminus\{d\}$ and, consequently, $Y_{x_H}\subset X_{x_H}$ according to statement \ref{aaa}. Since $X$ is individually rational, $x\in C_{x_H}(X_{x_H})$. Then, by substitutability, $x\in C_{x_H}(Y_{x_H}\cup\{x\})$.  This, together with our assumption that  $\{x\} P'_d  \{y\}$, implies that $x$ is a blocking contract for $Y$, leading to a contradiction. 

Therefore, statement \ref{cc} holds. Consequently,  
\begin{equation} \label{aa}
y\notin C_{y_H}(X_{y_H} \cup \{y\})
\end{equation} 
because otherwise, $y$ would block $X$ under $\widetilde{P}$.

Since $Y$ is individually rational, we have $y\in C_{y_H}(Y_{y_H})$. By \ref{aa} and the fact that $Y_{y_H}\subseteq X_{y_H}\cup\{y\}$ due to statement \ref{aaa},  there must exist a contract $z\in X_{y_H}\setminus Y_{y_H}$ such that $z\in C_{y_H}(X_{y_H} \cup \{y\})$. By substitutability, this implies 
\begin{equation}
    z\in C_{y_H}(Y_{y_H} \cup \{z\})
\end{equation}

Furthermore, since $z\in X_{y_H}\setminus Y_{y_H}$, we have $z\neq y$ and therefore $z_D\neq d$. Consequently,
$\widetilde{P}_{z_D}:=\{z\},\varnothing $. This implies that $Y_{z_D}=\varnothing$ because $z\notin Y$. 
Thus, $z$ is a blocking contract for $Y$ under $\widetilde{P}$, leading to a contradiction. 

The contradiction arises from the assumption that a stable allocation $Y$ exists under $\widetilde{P}$,  such as $Y\neq X$. Thus, $X$ is the only stable allocation under $\widetilde{P}$. 

To prove the second equality, we need to show that if   $X'$ is individually rational under $\widetilde{P}$ and $X_i'\widetilde{R}_iX_i$ for all $i\in D$, then $X'=X$

In fact, by the definition of $\widetilde{P}_{-d}$, it follows that $X_i\widetilde{R}_iX_i'$ for each $i\in D\setminus\{d\}$. Then, \begin{equation}\label{aab}
X_i=X_i' \text{ for all } i\in D\setminus\{d\}.
\end{equation}

Now, suppose that $X'_d  P'_d  X_d  R'_d \emptyset $. Then,  there exists a contract $w\notin X$ such that $X'_d=\{w\}$ and, consequently,  $w\in C_d( X\cup\{w\})$. Since $w\notin X$ and there is no contract in $\mathbf{X}$ involving both $w_H$ and $d$ other than $w$, we have  $X_{w_H}\subseteq \cup_{i\in D\setminus\{d\}} X_i=\cup_{i\in D\setminus\{d\}} X'_i$, where the last equality follows from  \ref{aab}. Thus, $X_{w_H}\cup\{w\} \subseteq X'$. Moreover, since $X'$ is individually rational, $w\in C_{w_H}(X')$. By substitutability, we have $w\in C_{w_H}(X\cup\{w\})$. Hence, $w$ is a blocking contract for $X$ under $\widetilde{P}$, which is a contradiction. This completes the proof of our Claim. 

Finally, from \ref{ww}, \ref{ww1}  and Claim,  it follows that 
$W(P_{d},O^{\phi' }(P_{d}))R_{d}W(P_{d}O^{\phi' }(P'_{d}))$ for each $P_{d}^{\prime }\in \mathcal{P}_{d}$.
Hence, $\phi'$ is NOM.
\end{proof}
\vspace{20pt}

The next theorem states that $HDAM$ is NOM. The proof can be outlined as follows. Given a doctor $d$, consider an allocation produced by $HDAM$ that gives $d$ the worst possible outcome it could receive by truthfully reporting its preferences. Then, define a sub-profile of preferences where each of the remaining doctors declares the outcome they received in this allocation as the only acceptable one. The key claim of the proof is that, under this sub-profile, the allocation obtained by doctor $d$ through any deviation is not better than the outcome received by truthfully reporting its preferences. Therefore, the worst outcome under $HDAM$ at true preferences is at least as good as the worst outcome under $HDAM$ at any deviation.

\begin{theorem}
\label{h-optimal nom} Assume that for each pair $(d,h)\in D\times H$ there
is at most one contract $x\in \mathbf{X}$ such that $x_{D}=d$ and $x_{H}=h$, and that all agents have substitutable preferences. Then,  $HDAM$ is NOM.
\end{theorem}

\begin{proof} 
Given $d\in D$ and $P_{d}\in \mathcal{P}_{d}$, let $\widehat{X}$ be an allocation produced by $HDAM$ that assigns $d$ with the worst outcome it can receive when truthfully reporting its preferences, i.e., $\widehat{X}_d=W(P_{d},O^{HDAM}(P_{d})).$ Since $\mathbf{X}$ is finite, there exists a profile $%
\widehat{P}_{-d}\in \mathcal{P}_{-d}$ such that $%
HDAM_{d}( P_{d},\widehat{P}_{-d}) =\widehat{X}$.  
Consider the profile   $\widetilde{P}_{-d}\in \mathcal{P}_{-d}$ defined as follows:

$\widetilde{P}_{i}=\widehat{X}%
_{i},\varnothing .$   for each $i\in
D\backslash \{ d\}$
($\widetilde{P}_{i}:=\varnothing $  in case $\widehat{X} _{i}=\varnothing $)

Next, we will present a claim regarding this profile, which will be instrumental in proving the theorem.

\underline{Claim:} $\widehat{X}_{d} R_{d} HDAM (
P_{d}^{\prime },\widetilde{P}_{-d})$  for all $P_{d}^{\prime }\in 
\mathcal{P}_{d}\setminus \{ P_{d}\}$.

Assume, on contradiction, that there exists $P_{d}^{\prime }\in \mathcal{P}%
_{d}\setminus \{ P_{d}\} $ such that 
$
HDAM_{d}(
P_{d}^{\prime },\widetilde{P}_{-d}) P_{d}\widehat{X}_{d}. $

Then, $HDAM_{d}(P_{d}^{\prime },\widetilde{P}_{-d}) =\{ z\} $ for some contract $z\in\bold{X}_d$,  since $HDAM$ is individually rational. Therefore, \begin{equation}
\{z\} P_{d}\widehat{X}_{d}R_{d}\varnothing.  \label{1a}
\end{equation}
Observe that $\widehat{X}_{z_H}\subseteq\cup_{i\in D\setminus\{d\}} \widehat X_i$ because $z$ is the only contract in $ \mathbf{X}$ involving both $d$ and $z_H$, and $z\notin\widehat{X}$.  Hence,  by definition of $
\widetilde{P}_{-d}$, the contracts in $\widehat{X}_{z_{H}}$ are never rejected by doctors along the HDA algorithm at $(
P_{d}^{\prime },\widetilde{P}_{-d}) $ (note that $\widetilde{P}_{x_{D}}:=\{x\},\varnothing $ for each $x\in 
\widehat{X}_{z_{H}}$). Let $T^{\prime }$ be the total number of iterations required for the HDA algorithm to converge under $(
P_{d}^{\prime },\widetilde{P}_{-d}) $ and let $X^{T^{\prime }-1}$ denote the set of contracts that have not been rejected until the end of stage  $T^{\prime }-1$, as defined in
HDA algorithm.  Then,  $\widehat{X}%
_{z_{H}}\subseteq X^{T^{\prime }-1}$ and, since $HDAM_{d}(
P_{d}^{\prime },\widetilde{P}_{-d}) =\{ z\}$, it follows that  $z\in
C_{H}(X^{T^{\prime }-1})$. Therefore, due to substitutability,  we have 
\begin{equation}
z\in C_{z_{H}}(\{z\}\cup \widehat{X}_{z_{H}})=C_{z_{H}}(\{z\}\cup \widehat{X})
\label{1b}
\end{equation}%
But (\ref{1a}) and (\ref{1b}) contradict the stability of $HDAM%
( P_{d},\widehat{P}_{-d}) $ . This concludes the proof of our Claim.

Therefore, for all $P_{d}^{\prime }\in 
\mathcal{P}_{d}^{\prime }\setminus \{ P_{d}\},$  the Claim implies
\begin{equation}W(P_{d},O^{%
HDAM }(P_{d})) =\widehat{X}_{d} R_{d} HDAM (
P_{d}^{\prime },\widetilde{P}_{-d})R_d W(P_{d},O^{HDAM }(P'_{d})). \end{equation} 
Thus, $HDAM$  is NOM.
\end{proof}

\begin{corollary}
\label{all nom} Assume that for each pair $(d,h)\in D\times H$ there
is at most one contract $x\in \mathbf{X}$ such that $x_{D}=d$ and $x_{H}=h$, and that all agents have substitutable preferences.
Then, both $EADAM$ and all stable-dominating mechanisms are NOM.
\end{corollary}
\begin{proof}
    The proof follows trivially from Remark \ref{duality} and  Theorems \ref{dominate NOM} and \ref{h-optimal nom}.
\end{proof}

\subsection {Matching with contracts}\label{without}

In this section, we consider markets where contracts hold significant importance. This means that $\mathbf{X}$  contains two or more contracts relating to the same pair of agents  $(d, h) \in \mathbf{D\times H}$, under different conditions. As a general hypothesis, we again assume that all agents have substitutable preferences.  Our first theorem shows that $DDAM$ remains NOM in these markets. To prove this result, we begin by stating the following lemma: at the end of each iteration of the DDA algorithm, each hospital is assigned to its choice set based solely on the offers it received during that iteration.
The proof of the theorem can be outlined as follows. Given a doctor $d$, we consider the worst outcome it can obtain through  $DDAM$ by truthfully declaring its preferences. If this outcome is not $d$'s first choice (otherwise, manipulations would not occur), we consider the allocation that $d$ ranks immediately below it, denoted as $\{x\}$. Utilizing the lemma, we identified a suitable sub-profile for the remaining doctors such that for any false preference $d$ could declare, it ends up with an outcome worse than $\{x\}$. This implies that such $DDAM$ is NOM.

\begin{lemma}
\label{L1} Given $P\in \mathcal{P}$, let $O^{t}$ and $O_{A}^{t}$ be definited as in DDA algorithm. Then, for every $h\in H$
\begin{equation}
C_{h}(O_{A}^{t})=C_{h}(O^{t})\text{ for all }t=1,...T  \label{ao}
\end{equation}%
 where $T$ is the number of iterations of the DDA algorithm at $P.$
\end{lemma}

\noindent \begin{proof} Given $P\in \mathcal{P}$, let $t$ be a stage of the DDA algorithm at $P$, such that $1\leq
t\leq T$ . We will prove the statement by induction on $t$. If $t=1$ the
proof is trivial. Now, assume that equation (\ref{ao}) holds for all $k<t$ and we will show it also holds for $t.$ 
By substitutability and induction hypothesis, $$C_{h}(O_{A}^{t})=C_{h}(O_{A}^{t-1}\cup O^{t})=C_{h} (C_{h}(O_{A}^{t-1})\cup O^{t})=C_{h}(C_{h}(O^{t-1})\cup O^{t}) $$
Next, we need to prove that $C_{h} ( O^{t-1})\subseteq O^t$ which will concludes the proof. Let $x\in C_{h} ( O^{t-1})$. Then $x\in O^{t-1}=C_{D}(X^{t-1})$ and consequently, $x\in X^{t-1}$. By the definition of the DDA algorithm and the induction hypothesis, it follows that $X^{t}=X^{t-1}\setminus(O^{t-1}\setminus C_H(O^{t-1})).$ Therefore, $x\in X^{t}$. By substitutability, $x\in C_{D}(X^{t})=O^{t}.$ Hence, $%
C_{h} ( O^{t-1}) \subseteq O^{t}.$ 

This completes the proof of the lemma.
\end{proof}

\begin{theorem} 
\label{d-optimal} Assume that all agents have substitutable preferences in a many-to-one matching model with contracts.
Then, $DDAM$ is NOM.
\end{theorem}

\begin{proof} Givem $d\in D$, $P_{d}\in \mathcal{P}_{d}$, let $\widehat{X}$ be an allocation produced by $DDAM$ that assigns to $d$ the worst possible outcome obtainable by truthfully reporting its preferences, i.e.,  $\widehat{X}_d=W(P_{d},O^{DDAM}(P_{d})).$ Since $\mathbf{X}$ is finite, there exists a profile $%
\widehat{P}_{-d}\in \mathcal{P}_{-d}$ such that $%
DDAM( P_{d},\widehat{P}_{-d}) =\widehat{X}$.

If $\widehat{X}_d$ were $d$'s first-ranked option, then  $DDAM$ would not admit manipulations by $d$ at $P_d$.
Otherwise, let $\{ x\} $ be the option that $d$ ranks lowest among those it considers preferable to $\widehat{X}_d$,  i.e.,  
$\{ x\} =\min\limits_{P_d} \{ \{ w\}\subseteq \mathbf{X}%
_{d}:\{ w\} P_{d} \widehat{X}_d \} $. Through the Claim below, we will show that for each false preference that $d$ might declare, the worst outcome it could receive will be ranked lower than $\{ x\} $ according to $P_d$. Consequently, it cannot be ranked higher than  $\widehat{X}_d$. Therefore,   \begin{equation}
W_{d}(P_{d},O^{%
DDAM}(P_{d}))\text{  }   R_{d} \text{  }W_{d}(P_{d},O^{DDAM}(P_{d}^{\prime })) \text{  for all } P_{d}^{\prime }\in \mathcal{P}_{d}\setminus \{ P_{d}\} .
\end{equation}
This means that $\overline{%
\phi }$ is NOM, as we aimed to prove. 

\underline{Claim:} $\{
x\} $ $P_{d}$ $W_{d}(P_{d},O^{DDAM}(P_{d}^{\prime }))$ for
all $P_{d}^{\prime }\in \mathcal{P}_{d}\setminus \{ P_{d}\} .$%

We will identify a specific sub-profile of preferences $\widetilde{P}_{-d}\in \mathcal{P}_{-d}$ that leads $d$ to receive an outcome worse than $\{ x\} $ when  declaring any false preferences $P_{d}^{\prime }.$

Let $T$ be the total number of iterations that DDA algorithm requires to converge at $(P_{d},\widehat{P}_{-d})$.  
For each $t=1,\ldots, T$, let $%
O^{t}$ be the set of all offers made by doctors in Stage t of DDA algorithm at $ (P_{d},\widehat{P}_{-d}) $ and set $O_{A}^ t:=\cup_{k=1}^{t}\overline O^{k}.$

Since $\{ x\} P_{d}\widehat{X}_d =DDAM_{d} (
P_{d},\widehat{P}_{-d}),$ there exists $\overline k\in \{ 1,...,T\} $
such that  
\begin{equation}
    x\in O_A^{\overline k}
\end{equation}
and
\begin{equation}\label{10}
    x\notin C_{x_{H}} ( O_A^{\overline k}), 
\end{equation}
which indicates that there exists a stage of DDA algorithm at $ (
P_{d},\widehat{P}_{-d}) $ where $x$ is offered by $d$ and rejected by $x_{H}$.  

Now, consider the specific profile of preferences $\widetilde{P}_{-d}\in \mathcal{P}_{-d}$  defined as follows:

$\widetilde{P}_{i}:=[C_{H} (
O^{\overline k}) ]_{i} ,\varnothing $   for each $i\in
D\backslash \{ d\}$
($\widetilde{P}_{i}:=\varnothing $  in case $[C_{H} (
O^{\overline k}) ]_{i}=\varnothing $). 

We will prove that $\{ x\} P_{d}%
DDAM_{d}  ( P_{d}^{\prime },\widetilde{P}_{-d}) $ for
all $P_{d}^{\prime }\in \mathcal{P}_{d}\setminus \{ P_{d}\} .$

Assume, for the sake of contradiction, that there exists $P_{d}^{\prime }\in \mathcal{P}_{d}\setminus
\{ P_{d}\} $ such that $DDAM_{d} ( P_{d}^{\prime
},\widetilde{P}_{-d}) =\{ z\} $ and $\{ z\}
R_{d}\{ x\} .$ Let $\overline T$ be the number of iterations of
the DDA algorithm at $ ( P_{d}^{\prime },\widetilde{P}_{-d}).$ W.l.o.g. we
can assume that $\overline T=T$ (if this is not the case, we can add some artificial
stages at $ ( P_{d},\widehat{P}_{-d})$ or $ ( P_{d}^{\prime },\widetilde{P}_{-d})$, where each doctor offers again the contract assigned by the DDA algorithm in the profile). 
For all $t=1,\ldots, T$, let $
\overline O^{t}$ be the set of all offers made by doctors in Stage t of DDA algorithm at $ (P_{d}^{\prime },\widetilde{P}_{-d}) $ and set $\overline O_{A}^ t:=\cup_{k=1}^{t}\overline O^{k}.$

Since $\{ z\} R_{d}\{ x\}
P_{d}\widehat{X}_d R_{d}\varnothing $, we have  $\{ z\}\neq \varnothing $ and $%
z_{H}\in H$ is well defined. Now, by definition of DDA algorithm,

\begin{equation}
z\in C_{z_{H}} ( [\overline O_{A}^{T}])   \label{7}
\end{equation}
and, due to substitutability, 
\begin{equation}
C_{z_{H}} ( [\overline O_{A}^{T}]_d) =\{ z\}  \label{07}
\end{equation}

By definition of $\widetilde{P}_{-d},$ $[C_{H} (
O^{\overline k}) ]_{i}=[\overline O^{1}]_{i}=[\overline O_{A}^{T}]_{i} $ for each $i\in D\backslash \{d\}$. In fact, every $i\in D\backslash \{d\}$ offers at most one contract along the DDA algorithm  at $ (
P_{d}^{\prime },\widetilde{P}_{-d}) $, $[C_{H} (
O^{\overline k}) ]_{i}$, starting from the first iteration. By (\ref{10}) and Lemma \ref{L1}, $x\notin C_{x_{H}} ( O^{\overline k}) $.
Since $d$ makes at most one offer per stage of the algorithm, it remains unmatched at the end of the iteration $\overline k$, i.e., %
\begin{equation}
\lbrack C_{H} ( O^{\overline k}) ]_{d}=\varnothing.  \label{5}
\end{equation}

Then, 
\begin{equation} 
C_{H}( O^{\overline k})=(\cup_{i\in D\backslash
\{d\}}[C_{H}( O^{\overline k}) ]_{i})\cup [C_{H} ( O^{\overline k}) ]_{d}=  \cup_{i\in D\backslash \{d\}}[\overline O_{A}^{
T}]_{i} 
\end{equation}
 
As a consequence, by applying properties of the choice sets, we obtain
\begin{equation}  C_{z_{H}}( O^{\overline k})=C_{z_{H}}( C_{z_{H}}( O^{\overline k}))=C_{z_{H}}( C_{H}( O^{\overline k}))=C_{z_{H}}( \cup_{i\in D\backslash \{d\}}[\overline O_{A}^{
T}]_{i})\label{80}\end{equation}

Then, again by properties of the choice sets, we can write: 
$$%
C_{z_{H}}( \overline O_{A}^{ T}) =C_{z_{H}}(
[\overline O_{A}^{T}]_{d} \cup ( \cup_{i\in
D\backslash \{d\}}[\overline O_{A}^{T}]_{i} )) =C_{z_{H}}( C_{z_{H}}(
[\overline O_{A}^{T}]_{d}) \cup C_{z_{H}}( \cup_{i\in
D\backslash \{d\}}[\overline O_{A}^{T}]_{i}) ) .$$

Hence, by (\ref{07})
and (\ref{80}), 
$$C_{z_{H}}( \overline O_{A}^{T}) =C_{z_{H}}(
\{ z\} \cup C_{z_{H}}( O^{\overline{k}}) ) .$$

Consequently, by (\ref {7}) and Lemma \ref{L1},
\begin{equation}
z\in C_{z_{H}}(
\{ z\} \cup C_{z_{H}}( O_A^{\overline{k}}) ).  \label{3}
\end{equation}

Given that $\{ z\} R_{d}\{ x\} $ and $x\in O^{ \overline{k}}$, $z$ was offered by $d$ before Stage $\overline {k}$. Then, $z\in
O_{A}^{\overline k}$. Hence, by (\ref{3}), $z\in C_{z_{H}}(O_A^{\overline{k}})$. This contradicts (\ref{5}) according to Lemma \ref{L1}.

This concludes the proof of our Claim, and Theorem \ref%
{d-optimal} follows. 
\end{proof}
\vspace{20pt}

Surprisingly, the $EADAM$ and $HDAM$ become obviously manipulable, even in the simplest case of the one-to-one matching model with contracts.\footnote{%
The one-to-one matching model can be viewed as a particular case of the many-to-one model where only singleton sets of contracts are acceptable for every  $h\in H$.} 
In the case of $EADAM$, a doctor $d$ may falsely declare a particular “unacceptable” contract with a hospital as “acceptable” which can lead to a contract between another doctor $d'$ and the same hospital to become an interrupting contract. Consequently, this interrupting contract must be removed from $d'$'s preference. As a result, a contract that is more preferred by $d$ could be definitively accepted by the corresponding hospital in an early step of the DDA algorithm, since  $d'$ no longer offers the removed contract. We provide an example demonstrating how such manipulations are possible. Furthermore, in such example, the allocation selected by  $EADAM$ is the only efficient outcome that Pareto dominates $DDAM$. Consequently, any efficient mechanism that Pareto dominates $DDAM$ becomes obviously manipulable. 

In the case of $HDAM$, a doctor may falsely declare a particular “acceptable” contract as “unacceptable”  leading it to be rejected when proposed by the corresponding hospital. As a result, the hospital may offer more favorable terms to the same doctor in a subsequent round. This scenario differs from $DDAM$ where doctors can begin by proposing their preferred contracts. Manipulating $DDAM$ (which is possible for some substitutable preferences) requires initiating a more complex “rejection chain,” involving other doctors. 
As previously mentioned, manipulations of $EADAM$ and $HDAM$ could include only contracts between the manipulating doctor and the involved hospital, making these manipulations quite obvious. These kinds of manipulations always leave the manipulating doctor with an outcome at least as good as truth-telling \textit{regardless of} the preference profile of the other doctors. They seem quite straightforward and are likely to be understood by the agents involved. Therefore, we will refer to them as  \textit{very} obvious manipulations.

\begin{definition}
Let $\phi :\mathcal{P}\mathbf{\rightarrow }A(\mathbf{X})$ be a mechanism,  $%
d\in D$ with true preference  $P_{d}\in \mathcal{P}_{d}$, and let $P_{d}^{\prime }\in \mathcal{%
P}_{d}$ be a manipulation of $\phi $ at $P_{d}$. Then, $P_{d}^{\prime }$ is a \textbf{very obvious manipulation} if 
\begin{equation}
\phi (P_{d}^{\prime },P_{-d}) R_{d} \phi (P_{d},P_{-d})
\end{equation}%
for all $P_{-d}\in \mathcal{P}_{-d}.$

 A mechanism is \textbf{very obviously manipulable} if it admits
a very obvious manipulation.
\end{definition}

A very obvious manipulation is an obvious manipulation. Therefore, if a mechanism is very obviously manipulable it is also obviously manipulable.

\begin{theorem}\label{om HDAM}
Assume that all agents have substitutable preferences in a matching model with contracts.
Then, \textit{$HDAM$ } and any efficient mechanism that Pareto dominates $DDAM$ (including \textit{$EADAM$}) are very obviously manipulable, even in the one-to-one setting.
\end{theorem}

\begin{proof}
First, we prove that $HDAM$ is very obviously manipulable. It suffices to show a specific market where a doctor $d\in D$ has a very obvious manipulation for $EADAM$.

Let $%
D=\{d_{1},d_{2}\}$, $H=\{h\}$ and $\mathbf{X}=\{w,x,y,z\}$ be the sets of
doctors, hospitals, and contracts, respectively.   Note that $w_{H}=x_{H}=y_{H}=z_{H}=h$ and assume that $%
w_{D}=x_{D}=y_{D}=d_{1}$, $z_{D}=d_{2}$. Let $\succ _{h}:=\{w\},\{x\},\{z\},\{y\}$ be the preferences of the hospital $h$.
For clarity, we will denote the preferences of doctor $d_{i}$ as $P_i$. 
 Suppose that $d_1$ has true preference  $P_{1}:=\{x\},\{w\},\varnothing $ and  let $P_{2}$ be an arbitrary preference in $\mathcal{P}_{d_{2}}.$  

Following the HDA algorithm, hospital $h$ offers contract $w$ to doctor $d_1$, and this offer is accepted. After that, no further offers are made.  Thus, $HDAM_{d_{1}}(P_{1},P_{2})=\{w\}$. However, if $d_1$ declares $P'_1\in \mathcal{P}_1$ such that  $P_{1}^{\prime }:=\{x\},\varnothing$, then following the HDA algorithm, $h$ first offers contract $w$ to $d_{1}$, who rejects its. Subsequently,   $h$ offers its second-best contract, $x$, to $%
d_{1}$ and this offer is accepted. After that, no further offers are made.  Thus, $HDAM%
_{d_{1}}(P_{1}^{\prime },P_{2})=\{x\}.$ Therefore, $HDAM%
_{d_{1}}(P_{1}^{\prime },P_{2})=\{x\} P_{1} \{w\}=HDAM%
_{d_{1}}(P_{1},P_{2}),$ for any $P_{2}\in \mathcal{P}_{d_{2}}.$ Hence, $%
P_{1}^{\prime }$ is a very obvious manipulation of $HDAM$ at $%
P_{1}.$

 Now, we prove that  $EADAM$ is very obviously manipulable. Consider the simple market presented above.
 Suppose that $d_1$ has true preferences $P_1:=\{y\},\varnothing $ and $P_2= \{z\}$. Since there are no interrupting pairs under $(P_1,P_2)$, we have $EADAM(P_1,P_2)=DDAM(P_1,P_2)=\{z\}$. However, if $d_1$ declares $P'_1\in \mathcal{P}_1$ such that $P'_{1}:=\{y\},\{x\},\varnothing$ then the DDA algorithm works as follows under $(P'_1,P_2)$
 
\begin{table}[h]
\begin{center}
\begin{tabular}{| r | l |}
\hline & h  \\ \hline
Step 1 & $\cancel{y},z$ \\ \hline
Step 2 & $x,\cancel{z}$ \\ \hline
\end{tabular}
\end{center}
 and $z$ is an interrupting contract. Therefore, by eliminating $z$ from $P_2$ we get that $EADAM(P'_1,P_2)=\{y\}$ and $P'_1$ is a manipulation of $EADAM$ under $P_1$. Furthermore, if $P'_2= \varnothing$, then $EADAM_{d_1}(P'_1,P'_2)=\{y\}=EADAM_{d_1}(P_1,P'_2)$. Therefore, since $\mathcal{P}_2=\{P_2,P'_2\}$, we conclude that  $P'_1$ is a very obvious manipulation of $EADAM$ under $P_1$.
\end{table}

 Observe that the outcomes selected by $EADAM$ at $(P_1,P_2)$ and $(P'_1,P_2)$ are the only efficient allocations that Pareto dominates the allocation selected by $DDAM$. Thus, any efficient mechanism that Pareto dominates $DDAM$ will coincide with $EADAM$ for these profiles and thus becomes very obviously manipulable.
\end{proof}

From Theorem \ref{om HDAM}, it follows that none of the three results obtained without contracts hold for the model with contracts, even in the one-to-one scenario. This highlights a significant contrast and a substantial difference between the models with and without contracts in terms of the strategic behavior of agents.

\section{Final Remarks} \label{section final}

Table \ref{tabla caracterizaciones}
summarizes our main findings. Note that these results apply to both many-to-one (with substitutable preferences) and one-to-one settings. From the table, two intriguing open questions emerge: Are there other NOM stable-dominating mechanisms besides $DDAM$ in matching models with contracts? And, is efficiency compatible with NOM in this model? 

\begin{table}[h]
\small
  \begin{center}
\begin{tabular}{|c|c|c|c|c|c|}

\cline {1-5}
& \multicolumn{4}{|c|}{NOM (for doctors)} \\
\cline{2-5}

& \multicolumn{2}{|c|}{Without Contracts} & \multicolumn{2}{|c|}{With
Contracts} \\
\hline
\multicolumn{1}{|l|} {DDAM} & \multicolumn{2}{|c|}{$yes$} & \multicolumn{2}{|c|}{$yes$} 

\\
\hline
\multicolumn{1}{|l|} {HDAM} & 
\multicolumn{2}{|c|}{$yes$} & \multicolumn{2}{|c|}{$no$} 
\\ \hline

\multicolumn{1}{|l|} {EADAM} & 
\multicolumn{2}{|c|}{$yes$} & \multicolumn{2}{|c|}{$no$} 

\\ \hline

\multicolumn{1}{|l|} {Any stable-dominating mechanisms} & 
\multicolumn{2}{|c|}{$yes$} & \multicolumn{2}{|c|}{$no$} 
\\ \hline
\end{tabular}
\end{center}
\caption{\emph{Summary of  results.}} 
\label{tabla caracterizaciones}
\end{table}

Additionally,  open questions arise when we interchange the roles of doctors and hospitals and consider the manipulability of mechanisms by hospitals. Even in the college admission model, no stable and non-manipulable mechanism for colleges exists  \citep[see ][]{roth1985college}. Therefore, even in this straightforward context, the consideration of NOM (for colleges) mechanisms presents an interesting problem.

\bibliographystyle{ecta}
\bibliography{biblio}

\end{document}